\documentclass[%
 reprint,
 amsmath,amssymb,
 aps,
prab,
floatfix,
]{revtex4-1}

\usepackage[pdftex]{graphicx}
\usepackage{subfigure}
\usepackage{dcolumn}
\usepackage{bm}
\usepackage{dcolumn}
\usepackage{subfloat}
\usepackage{booktabs}
\usepackage{float}
\usepackage{epstopdf}
\usepackage{gensymb}

\begin{document}

\preprint{APS/123-QED}

\title{Multi-beam-energy operation for the continuous-wave x-ray free electron laser}

\author{Jiawei Yan$^{1,2}$} 
\author{Haixiao Deng$^{3,1}$}%
\email{denghaixiao@zjlab.org.cn}  \affiliation{%
	$^1$Shanghai Institute of Applied Physics, Chinese Academy of Sciences, Shanghai 201800, China\\ $^2$University of Chinese Academy of Sciences, Beijing 100049, China\\ $^3$Shanghai Advanced Research Institute, Chinese Academy of Sciences, Shanghai 201210, China. 
}%

\date{\today}
\begin{abstract}

The parallel operation of multiple undulator lines with a wide spectral range is an important way to increase the usability of x-ray free electron laser (XFEL) facilities, especially for machines with high-repetition-rate. In this paper, a delay system based on four double bend achromats is proposed to delay electron beams, thereby changing the arrival time of those delayed electron beams in the accelerating structure behind the system. Combined with kickers, the delay system can be used to generate bunch-to-bunch energy changed electron beams in a continuous wave XFEL facility. Start-to-end simulations based on the parameters of the Shanghai high-repetition-rate XFEL and extreme light facility are performed to demonstrate that the delay system can flexibly control electron beam energy from 1.48 to 8.74 GeV at the end of the linac.

\end{abstract}

\pacs{61.10.Ht,41.60.Cr, 07.85.Nc}
\maketitle

\section{Introduction}

X-ray free electron lasers (XFELs) play an important role in many research fields such as biology, nonlinear physics, and material science due to their high peak brightness, short pulse duration, and coherence \cite{Pellegrinireview, feng2018review}. However, the linear accelerator-based XFELs are unable to build as many experimental stations as the synchrotron radiation facility based on the storage ring, which makes it difficult to meet rapidly increasing user demands. Besides, different XFEL users may require different radiation wavelengths, thus the XFEL radiation is preferably obtained over a wide spectrum, especially covering soft x-ray to hard x-ray. 

To increase the number of experimental stations, the multi-undulator-line operation \cite{faatz2016simultaneous, hara2016pulse} has been applied to most XFEL facilities, including those in operation or under construction. In the multi-undulator-line operation, electron beams are alternately transferred into different undulator lines by fast switching kickers. According to the resonance condition \cite{6}: 
\begin{equation} 
\lambda  = \frac{{{\lambda _u}}}{{2{\gamma ^2}}}(1 + \frac{{{K^2}}}{2}), 
\end{equation} 
where $\lambda $ is the central wavelength of the XFEL radiation, ${\lambda _u}$ is the undulator period length, $\gamma $ is the mean Lorentz factor of the electrons, and $K$ is the undulator field parameter, the photon energy of each undulator line can be controlled independently through adjusting the undulator field parameter. However, the tunable range of the photon energy is limited by the tunable range of the undulator magnetic field strength. A larger adjustable range of the photon energy requires adjustment of the electron beam energy, which will affect the photon energy of all undulator lines. Based on the multi-undulator-line operation, two schemes have been proposed to generate XFEL radiation over a wide spectrum. The first method is to branch out the electron beams with low energy in the middle of the linac and transfer them into the undulator lines with low photon energies \cite{swissfel}. This method is mainly to generate soft x-ray and hard x-ray FEL in different undulator lines. It needs an additional bypass beam transport line. In addition, since the wavelengths of the XFEL pulses depend on two kinds of electron beams having a large difference in energy, it is difficult to continuously adjust the radiation wavelength over a wide range. Besides delivering low-energy electron beams into specific undulator lines, the SPring-8 Angstrom Compact free-electron LAser (SACLA) proposed and verified a scheme that can obtain multiple energy electron bunches by setting some specific rf units of the linac at the subharmonics of the trigger frequency \cite{hara2013time}. In this scheme, some electron beams will not be accelerated when passing through those specific rf units, which means that these electron beams will have lower energy at the exit of the linac. The advantage of this method is that the electron bunch energy can be controlled by changing the number of rf units with a specific trigger frequency and the corresponding trigger frequency without destroying the bunch quality. However, it can only be applied to those machines operated in a pulsed mode, not to the continuous-wave (CW) superconducting linac.

In recent years, the high-repetition-rate XFEL based on superconducting linac has received increasing attention due to the ability to produce radiation with higher average brightness. European XFEL \cite{european} achieved its first lasing and started the operational phase in 2017. The Linac Coherent Light Source II (LCLS-II) \cite{galayda2014linac} is currently under construction and is scheduled to be available to users in 2020. Further, the construction of the Shanghai High-Repetition-Rate XFEL and Extreme Light Facility (SHINE) \cite{zhu2017sclf} began in 2018. In this paper, a newly designed delay system is proposed to generate bunch-to-bunch electron beams with large energy differences in the CW superconducting linac. This system is applied to the SHINE and is located before the last accelerating structure of the SHINE linac.  The accelerating phase of those delayed electron beams in the last accelerating section is changed, so their energy is different from that of the electron beams without delay at the end of the accelerating structure. Numerical simulations show that this delay system can delay a 5.11 GeV electron beam up to 115 mm, thus flexibly controlling the electron beam energy from 1.48 to 8.74 GeV at the end of the linac.

\begin{figure*}[htp] 
	\centering 
	\includegraphics[width=\linewidth]{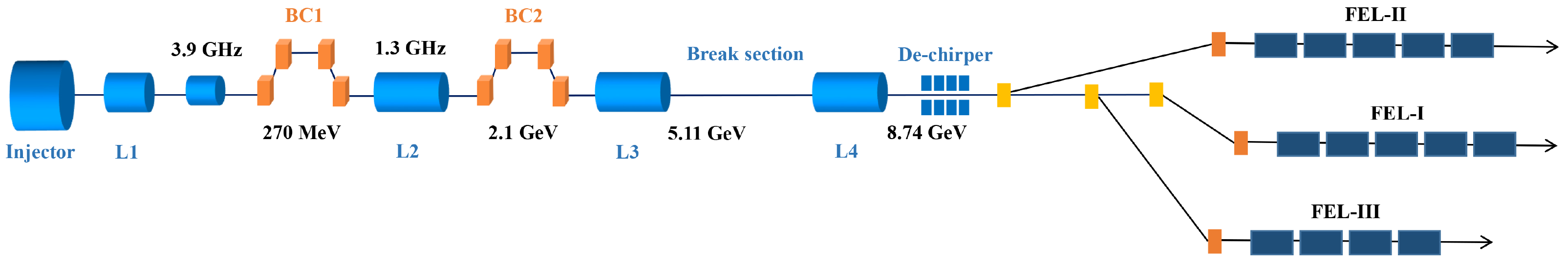}
	\caption{Schematic layout of the SHINE.}
	\label{FIG1} 
\end{figure*}

Since the numerical simulations are based on the SHINE parameters, this facility is briefly described in Sec.\ II. In Sec.\ III, the design of the delay system is introduced in detail. The start-to-end numerical simulation results are shown in Sec.\ IV. The conclusions and future applications of this delay system are summarized in Sec.\ V.

\section{Brief introduction to the SHINE}

As the first hard x-ray FEL in China, the SHINE is designed to deliver photons between 0.5 keV and 30 keV with a repetition rate up to 1 MHz. The schematic layout of the SHINE is shown in Fig.\ \ref{FIG1}. 

In the baseline design of the SHINE, 100 pC electron bunches with 100 MeV are generated in the photo-injector section, which includes a photocathode VHF gun, a 1.3 GHz buncher, a 1.3 GHz single 9-cell cavity cryomodule, a 1.3 GHz eight 9-cell cavities standard cryomodule, and a laser heater. Downstream of the photo-injector section is the main accelerator including four accelerating sections and two bunch compressors. Electron beams are accelerated at the off-crest phase of the first accelerating section (L1) with 2 cryomodules to create an energy chirp for the bunch compression. Following this, two 3.9 GHz cryomodules tuned at the 3rd harmonic of 1.3 GHz are used to linearize the energy chirp. The electron beams with an energy of 270 MeV are compressed in the first magnetic chicane (BC1). The second accelerating section (L2) with 18 cryomodules is located between the BC1 and the second magnetic chicane (BC2), which accelerates the electron beams from 270 MeV to 2.1 GeV and provides the additional energy chirp required for the bunch compression in the BC2. Behind the BC2, the electron beams are accelerated to 8.74 GeV at the on-crest phase of the third and fourth accelerating section (L3 and L4). There is a 150 m break section between the L3 and L4, which is prepared for future upgrades such as the third bunch compressor or an extraction line. Behind the L4, there is a corrugated structure \cite{dechirper} to compensate the correlated energy spread of the electron beam.

At the end of the main linac, the switchyard system distributes the electron beams to different undulator lines. The three undulator lines of the SHINE are referred to as the FEL-I, FEL-II, and FEL-III. Besides, spaces are reserved for the other three undulator lines in the future. All the undulators of the SHINE are chosen to be variable gap one. The baseline operation modes of the FEL-I and FEL-III are both self-amplified spontaneous emission (SASE) and hard x-ray self-seeding. Based on the normal planar undulator with a 26 mm period, FEL-I covers the photon energy of 3.58-17.90 keV. FEL-III covers the photon energy of 11.97-29.84 keV utilizing the superconductive undulator with a period of 16 mm. Based on the normal planar undulator with a 68 mm period, FEL-II covers the photon energy of 0.48-3.58 keV using external-seeding schemes and SASE.

In order to enable the radiation wavelength of each undulator line to be independently adjustable over a wider range, a delay system is considered to be placed between the L3 and L4 for the multi-beam-energy operation. Since the arrival time is changed, the electron beams passing through the delay system are at the off-crest phase of the L4. If an electron beam is delayed by a quarter of the rf period, it will be not accelerated in the L4, and it will be decelerated in the L4 when the delay distance is between one quarter and one half of the rf period. Therefore, the delay distance provided by this system is expected to be flexibly controlled from zero to one-half of the rf period to obtain the maximum electron beam energy control range. For the 1.3 GHz rf accelerating section, one-half of its rf period is 115.38 mm.

\section{DESIGN OF THE Electron beam delay system}

Magnetic chicane is widely used in XFELs for electron bunch delay \cite{zhang2015flexible}. The extra distance delay induced by a magnetic chicane \cite{dohlus2005bunch} is
\begin{equation}
\Delta Z = \left | \frac{R_{56}}{2} \right |.
\end{equation}
If a normal magnetic chicane is used to delay electron beams by 115 mm, its $\left | R_{56}  \right |$ is required to be 230 mm, which will destroy the quality of the electron beam with energy chirp. To minimize the impact on the electron beam quality, the delay system should be achromatic and isochronous. Therefore, a delay system based on the double bend achromat (DBA) is proposed for delaying electron bunches up to 115 mm. 

\begin{figure}[htp] 
	\centering 
	\includegraphics[width=\linewidth]{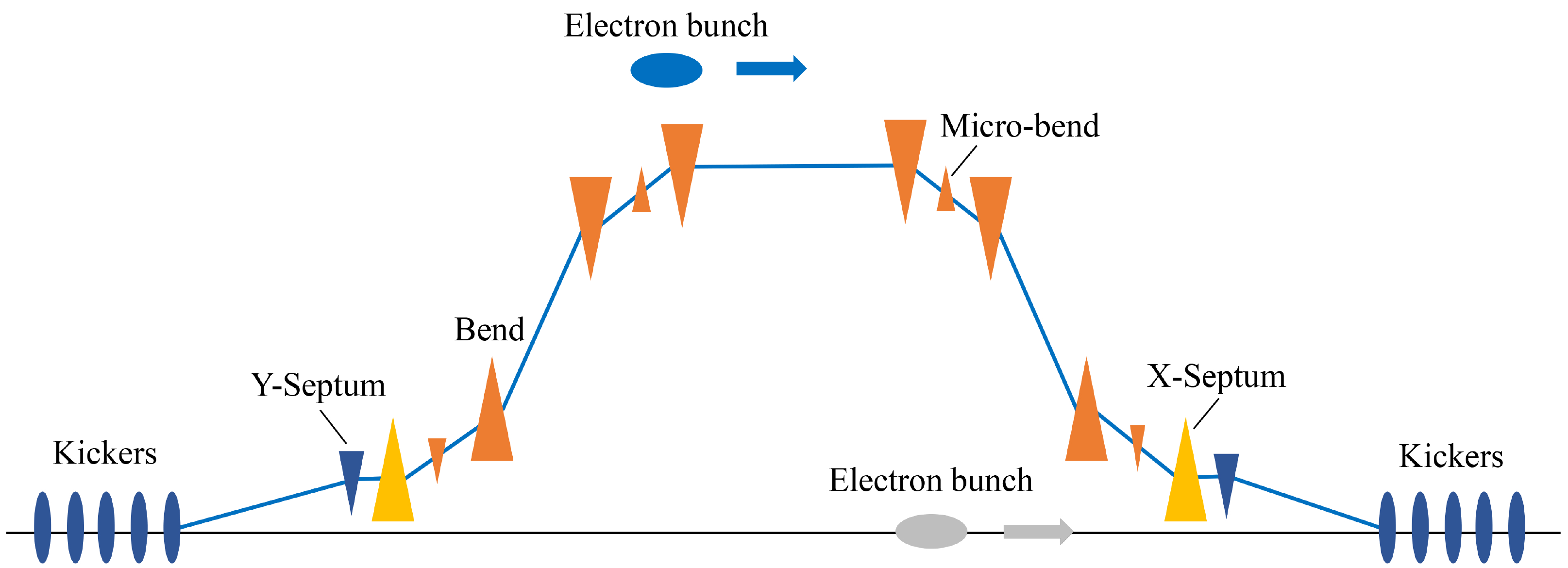}
	\caption{Schematic layout of the electron beam delay system and the kickers. Quadrupoles between the dipoles are not shown here.}
	\label{FIG2} 
\end{figure}

 The delay system consists of four horizontal DBAs. To make the lattice isochronous, a micro-bend \cite{milas2010design,chen2018design} is added in the mid-point of each DBA. The micro-bend has a small and inverse deflection angle compared with the corresponding dipoles of DBA. Due to the micro-bend, the $R_{56}$ can be decreased to below 1 $\mu m$ at the exit of the delay system. Besides, to separate the electron bunches that need to be delayed from the normally accelerated electron bunches, there are two sets of vertical kickers placed on both sides of the delay system. In this scheme, a set of kickers consists of five kickers with a 0.5 m length and a 0.2 mrad deflection angle. The distance between two adjacent kickers is 0.5 m. Behind the five kickers, there is a 15 m drift space to separate the kicked and un-kicked electron beams by 17.30 mm. At the end of the drift space, there is a small vertical septum with one vertical region and one field-free region to correct the deflection angle. Those un-kicked electron beams will pass the field-free region. Similarly, at the end of the delay system, there is a vertical septum provides a 1 mrad deflection angle to send the delayed electron bunches to the original orbit. Behind a 15 m drift space, there are five kickers to correct the deflection angle. In addition, due to the limited distance of the separation, the first dipole of the first DBA and the second dipole of the fourth DBA also employ septum magnets that with one horizontal region and one field-free region. The schematic layout of the delay system and the two sets of kickers are shown in Fig.\ \ref{FIG2}.  
 
 \begin{figure}[!htb]
 	\centering
 	\subfigure{\includegraphics*[width=0.85\linewidth]{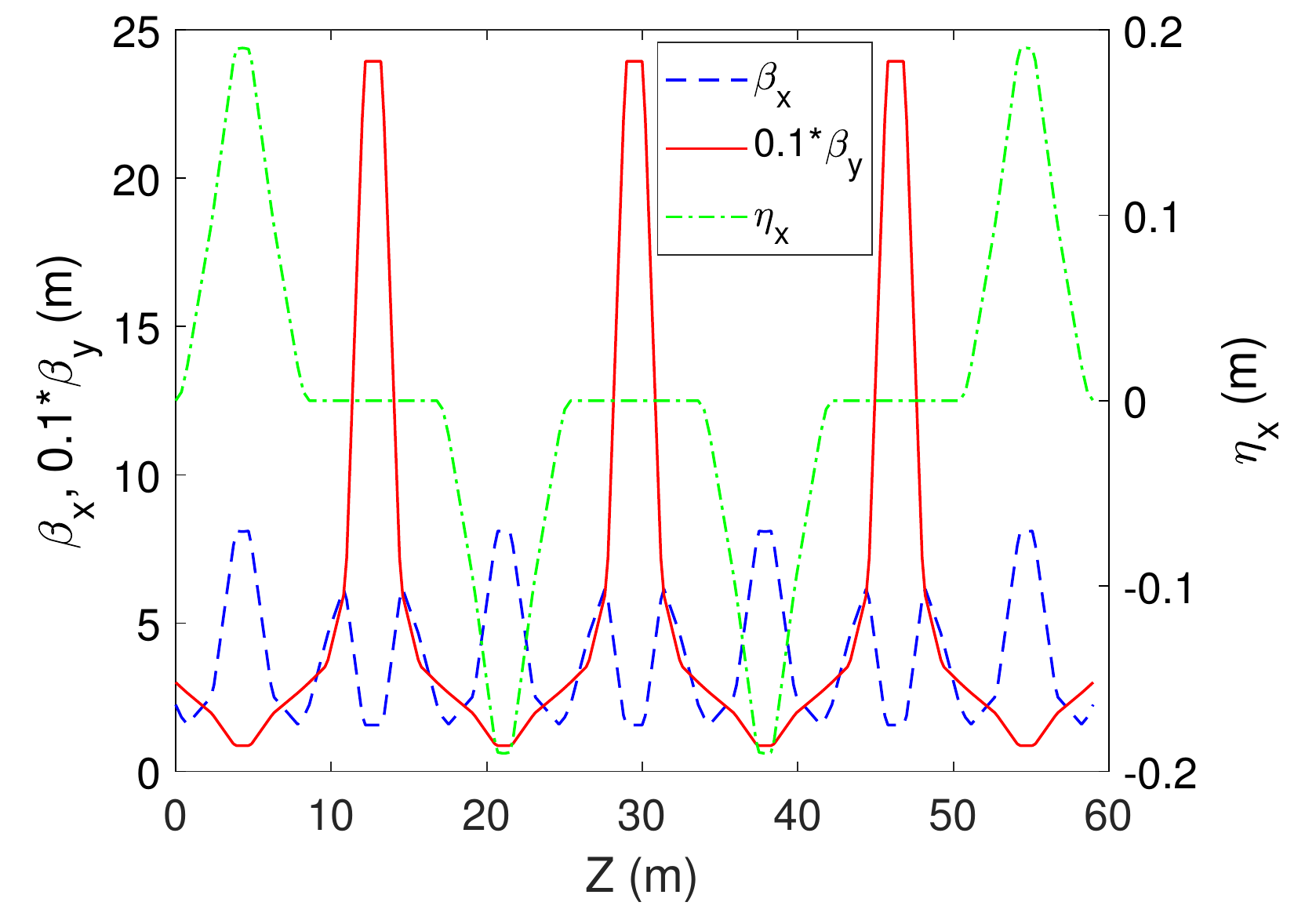}}
 	\subfigure{\includegraphics*[width=0.8\linewidth]{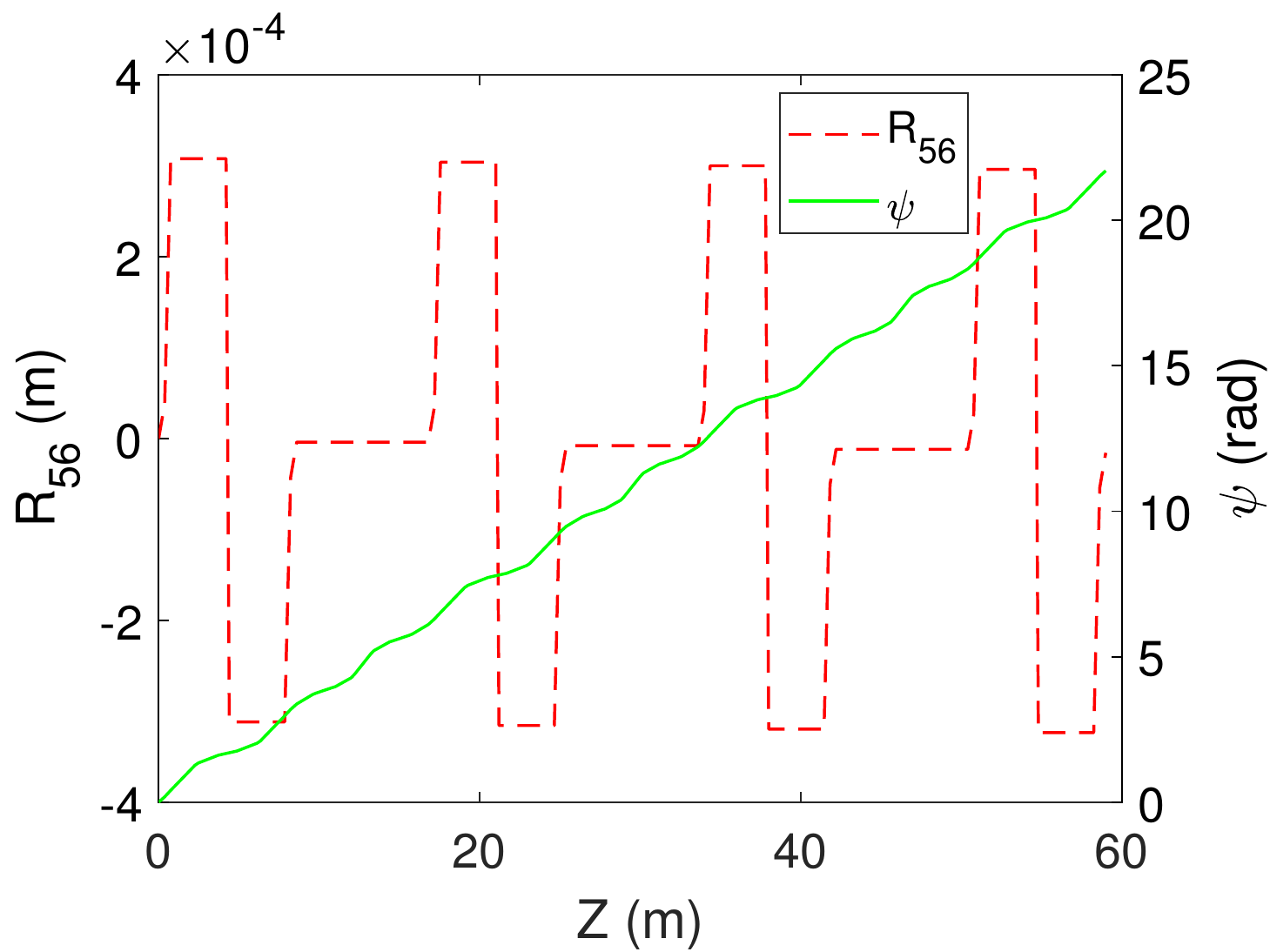}}
 	\caption{Upper: Courant–Snyder parameters $\beta_{x}$, $\beta_{y}$, and horizontal dispersion
 		function along the delay system. Bottom: $R_{56}$ and betatron phase advance in the horizontal plane along the delay system.}
 	\label{FIG3}
 \end{figure}
 
 The most critical issue of such DBA-based system is how to avoid the deterioration of electron bunch quality caused by the coherent synchrotron radiation (CSR) effect. To cancel the CSR kick, the optics balance method \cite{douglas1998thomas,di2013cancellation} is used in the lattice design of this delay system. The main idea of the optics balance method is to have successive CSR kicks separated by $\pi$ betatron phase advance in the bending plane, thus the slice transverse mismatch is canceled. To achieve this, there are four quadrupoles in each DBA and six quadrupoles in the drift space between two consecutive DBAs, controlling the phase advance between two consecutive dipoles to be $\pi$. In addition, the strength of these quadrupoles and the angle of the micro-bend also need to be optimized to make the system achromatic and isochronous. Therefore, the evolutionary algorithm \cite{yan2019generation} is used to optimize the lattice so that these requirements can be met simultaneously.
 
 The delay distance induced by one DBA can be estimated as:
 \begin{equation}
 \Delta Z_{DBA} = \frac{L_{1}}{2}  (\frac{1}{cos\left |\theta \right |}+\frac{1}{cos(\left |\theta \right |-\left |\theta_{m} \right |)}-2),
 \end{equation}
 where $L_{1}$ is the projection length of the DBA on the longitudinal axis, $\theta$ is the dipole angle of the DBA, and $\theta_{m}$ is the micro-bend angle. The delay distance induced by one inclined drift space between two consecutive DBAs can be calculated as:
 \begin{equation}
 \Delta Z_{D} = L_{2} (\frac{1}{cos(\left |2\theta \right |-\left |\theta_{m} \right |)}-1),
 \end{equation}
 where $L_{2}$ is the projection length of the drift space on the longitudinal axis. The delay distance induced by the whole delay system is:
  \begin{equation}
 \Delta Z = 4\Delta Z_{DBA}+2\Delta Z_{D}.
 \end{equation}
 More accurate delay distance can be obtained by ELEGANT \cite{elegant} simulation. In order to achieve a delay distance up to 115 mm, $L_{1}$, $L_{2}$, $\theta_{m}$, and $\theta$ are selected as 8.59 m, 8.16 m, 0.184\degree, and 2.819\degree, respectively. The delay distance can be well controlled by adjusting the dipole angle from 0\degree to 2.819\degree. The micro-bend angle needs to change correspondingly with the change of the dipole angle but has little effect on the delay distance. When the dipole angle is set as 2.819\degree, the $\beta$ functions, horizontal dispersion function, $R_{56}$, and horizontal phase advance $\psi$ of the optimized lattice along the system are presented in Fig.\ \ref{FIG3}.

\begin{figure}[!htb]
	\centering
	\subfigure{\includegraphics*[width=0.85\linewidth]{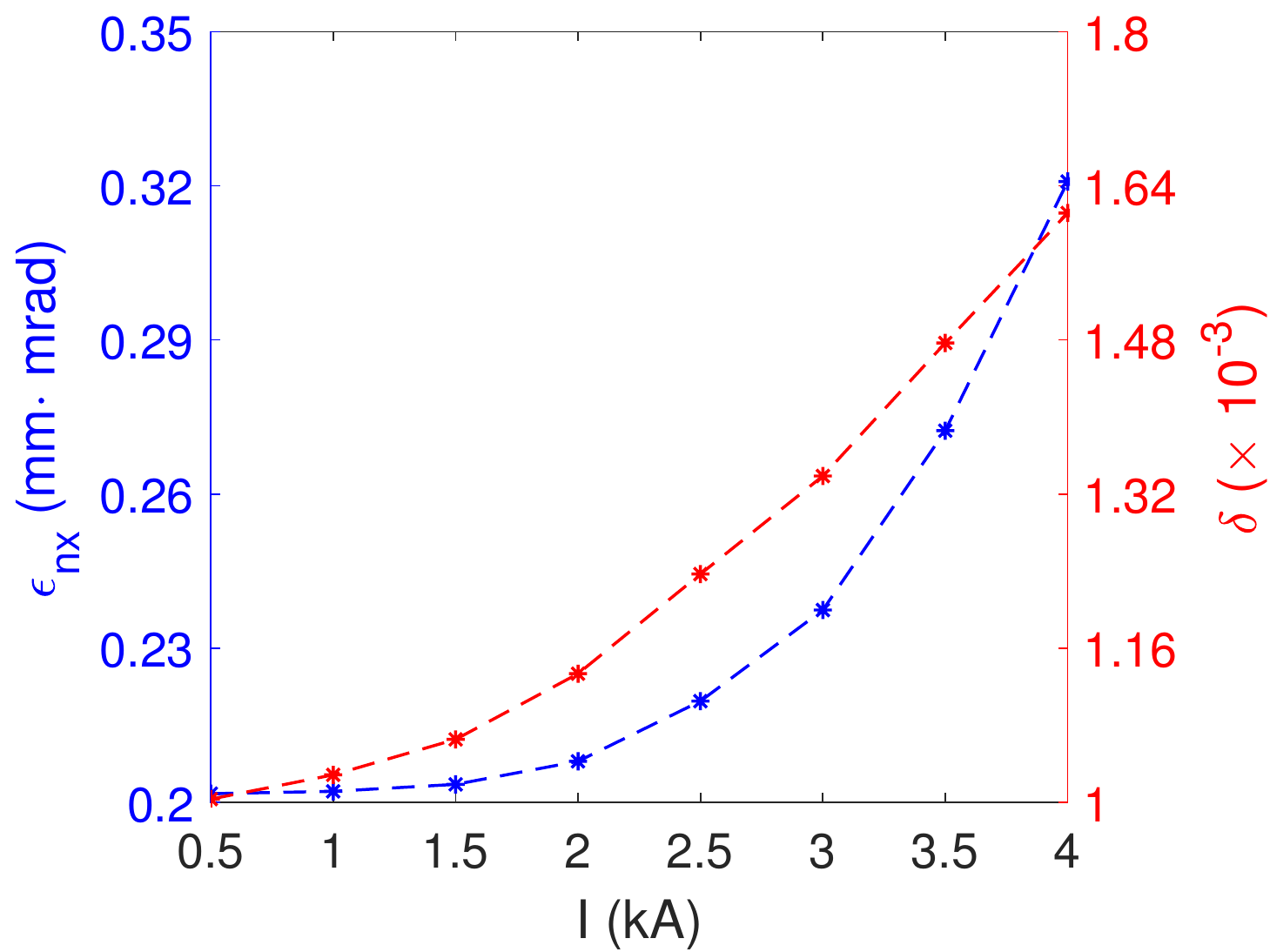}}
	\caption{The normalized horizontal emittance and energy spread of the Gaussian electron beams with different peak current at the end of the delay system.}
	\label{idealbeam}
\end{figure}

To further verify the optimized lattice, 100 pC Gaussian-distributed electron beams with different peak currents are sent into the delay system with a dipole angle of 2.819\degree. The energy spread and normalized transverse emittance of the electron beams at the entrance of the delay system are set to $1 \times 10^{-3}$ and $0.2 \rm mm \cdot mrad$, respectively. The tracking simulation is performed by ELEGANT with one million macroparticles, where the CSR effect is considered.  Fig.\ \ref{idealbeam} shows the energy spread and normalized horizontal emittance of these electron beams at the end of the delay system. The simulation results show that the growth of the horizontal emittance and energy spread induced by the delay system increases with the decrease of the bunch length and can be well controlled when the bunch length is in a reasonable range. 
 
\begin{figure}  
	
	\begin{minipage}[htbp]{0.5\linewidth}  
		\centering  
		\includegraphics[width=\columnwidth]{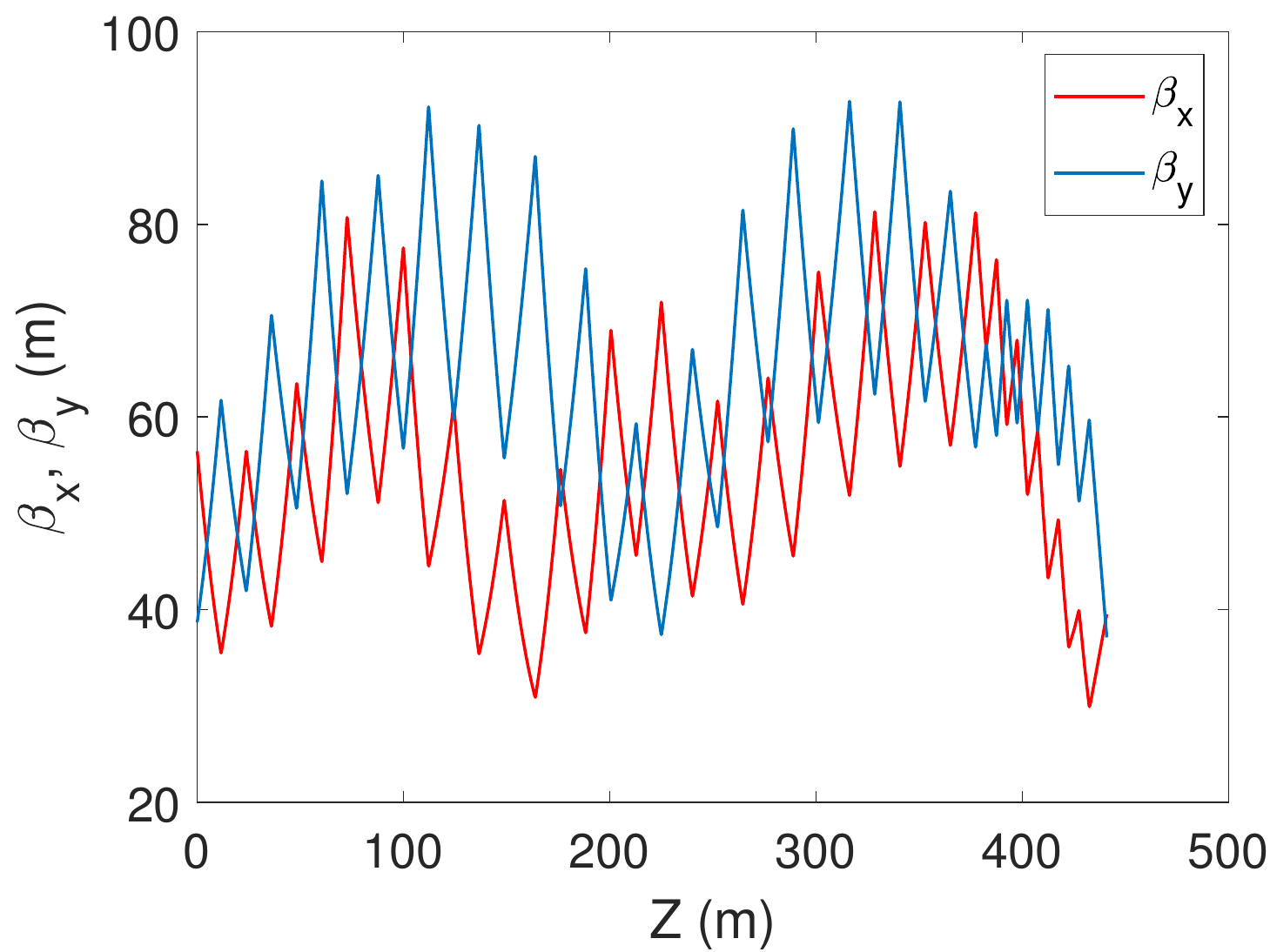}  
		
	\end{minipage}%
	\hfill 
	\begin{minipage}[htbp]{0.5\linewidth}  
		\centering  
		\includegraphics[width=\columnwidth]{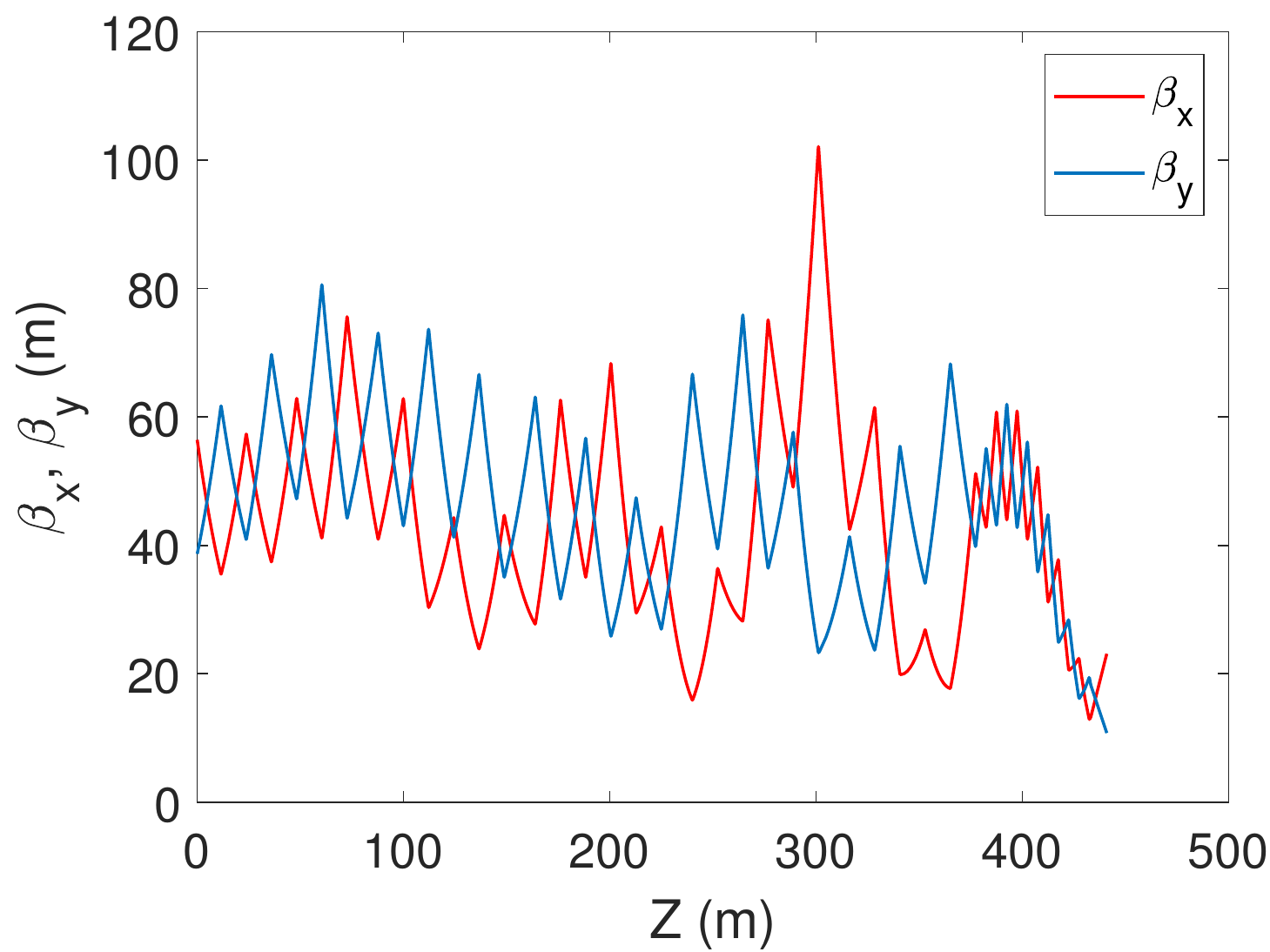}  
		
	\end{minipage}  
	\hfill 
	\begin{minipage}[htbp]{0.5\linewidth}  
		\centering  
		\includegraphics[width=\columnwidth]{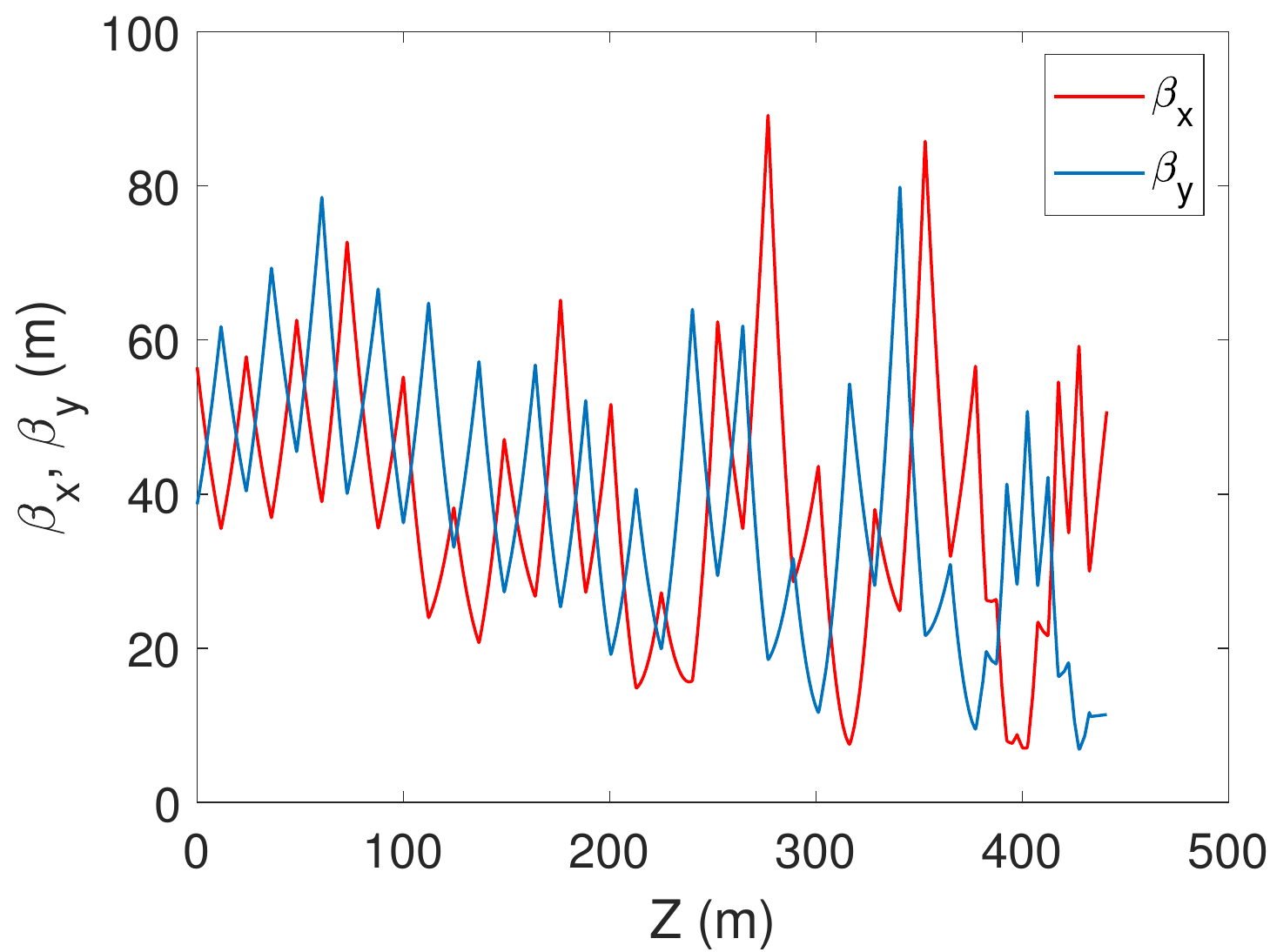}  
	\end{minipage}%
	\hfill 
	\begin{minipage}[htbp]{0.5\linewidth}  
		\centering  
		\includegraphics[width=\columnwidth]{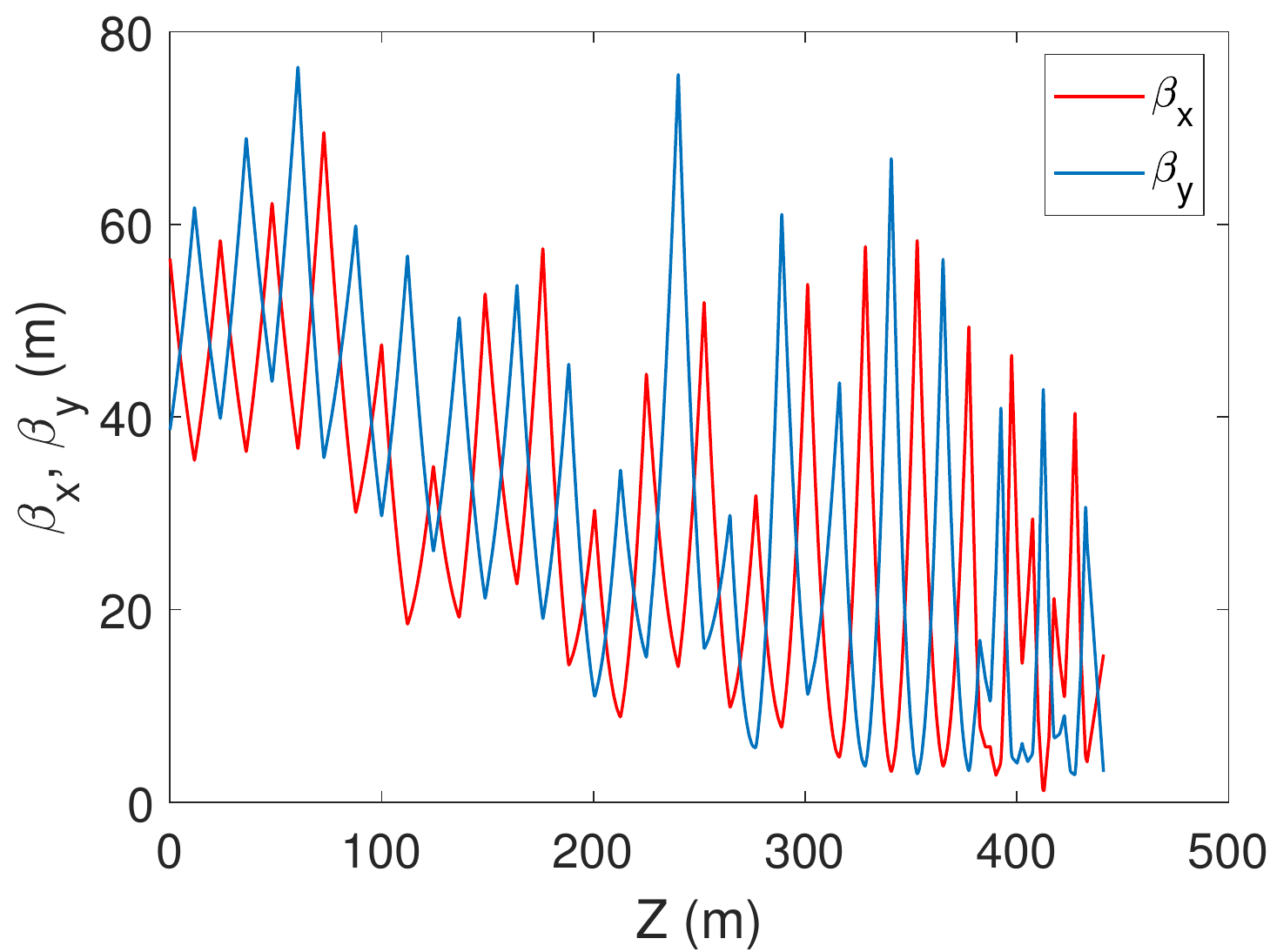}  
	\end{minipage}  
	
	\caption{The $\beta$ functions of the four electron bunches with typical energies of 8.74 (top left), 5.11 (top right), 3.30 (bottom left), and 1.48 GeV (bottom right), respectively, along the L4 and the dechirper section based on the optimized lattice.}
	\label{FIG4} 
\end{figure} 

\section{start-to-end simulation}

\begin{figure*}  
	\begin{minipage}[htbp]{0.25\linewidth}  
		\centering  
		\includegraphics[width=\columnwidth]{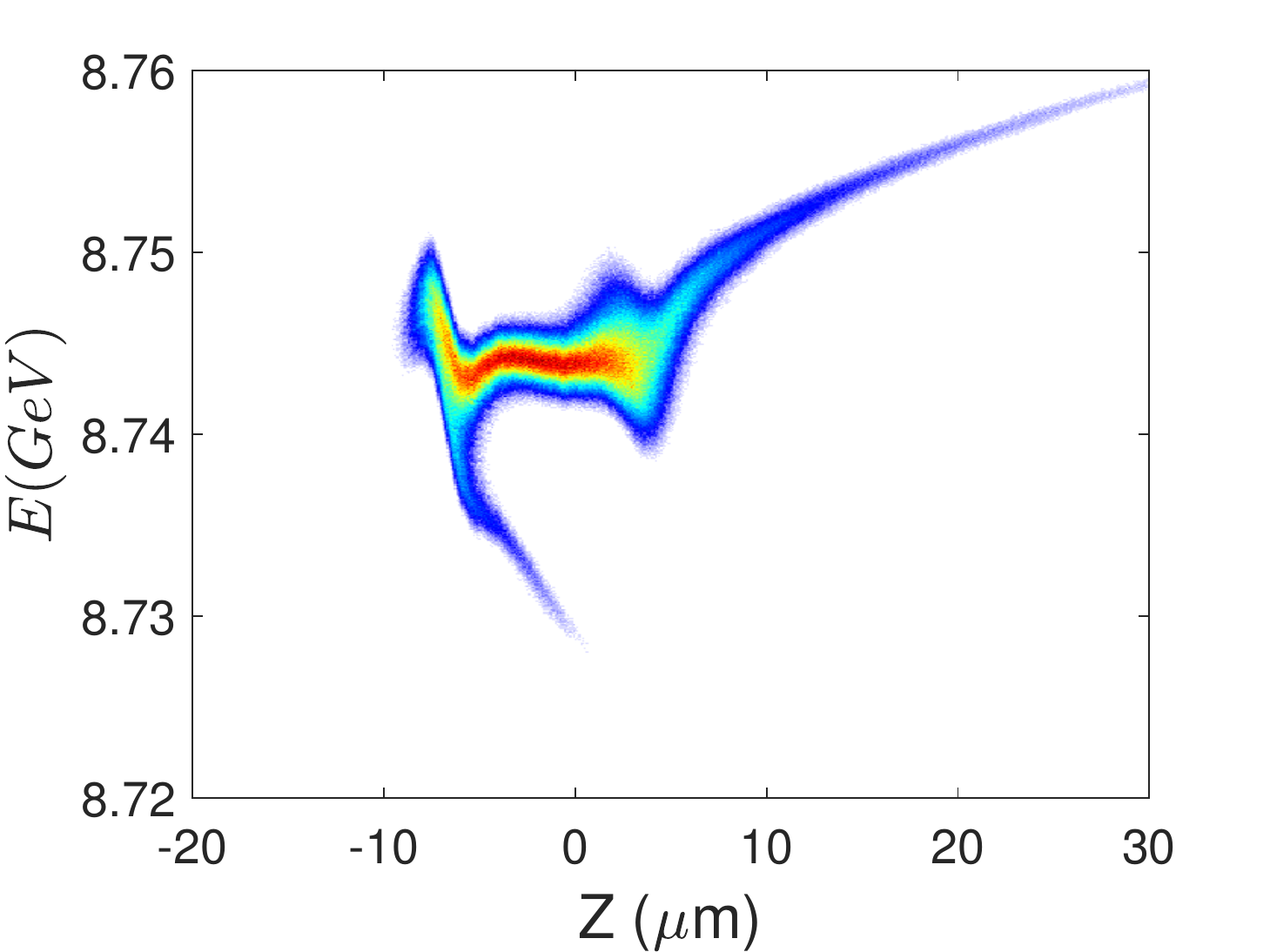}  
	\end{minipage}%
	\hfill 
	\begin{minipage}[htbp]{0.25\linewidth}  
		\centering  
		\includegraphics[width=\columnwidth]{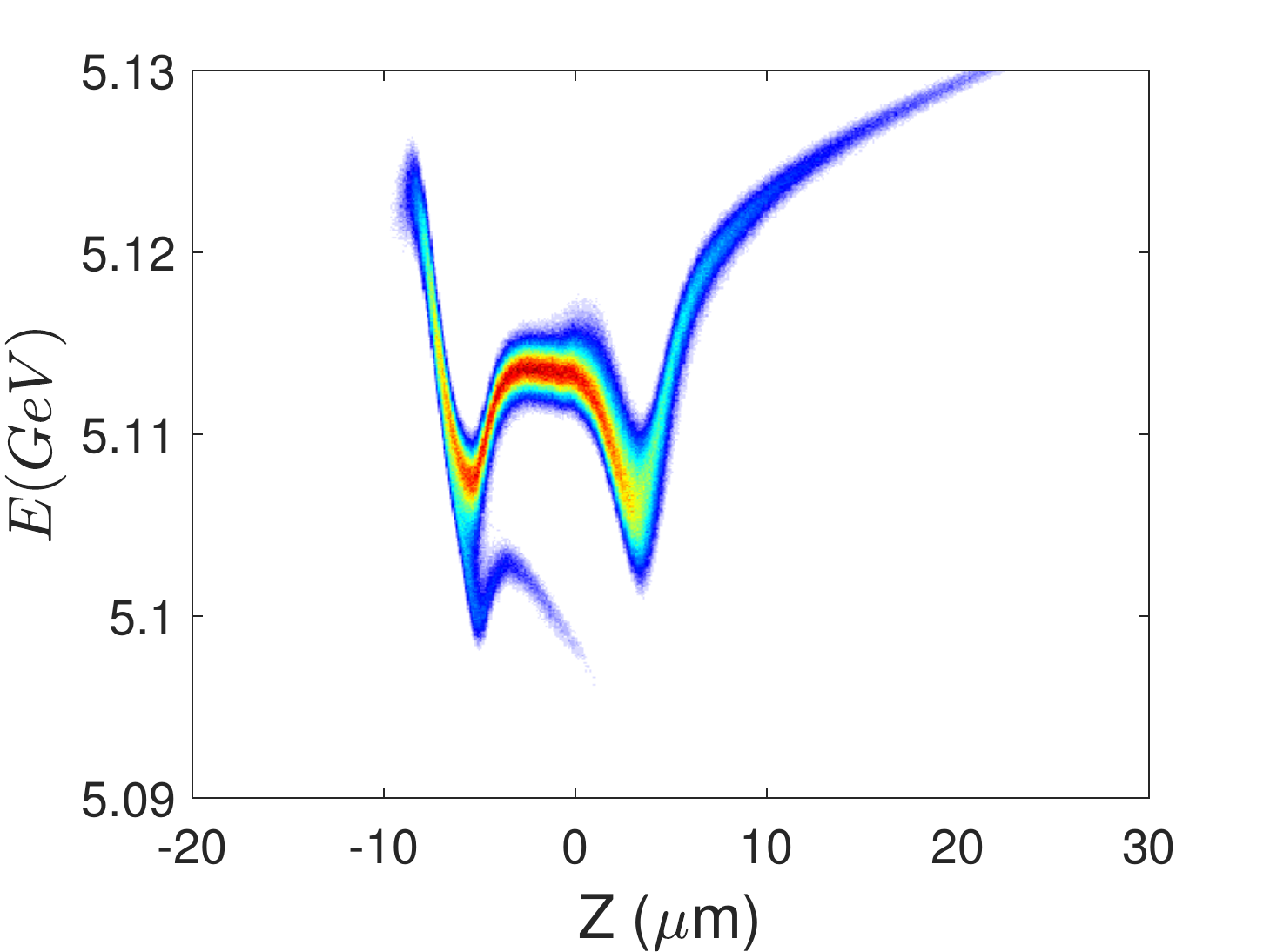}  
	\end{minipage}%
	\hfill 
	\begin{minipage}[htbp]{0.25\linewidth}  
		\centering  
		\includegraphics[width=\columnwidth]{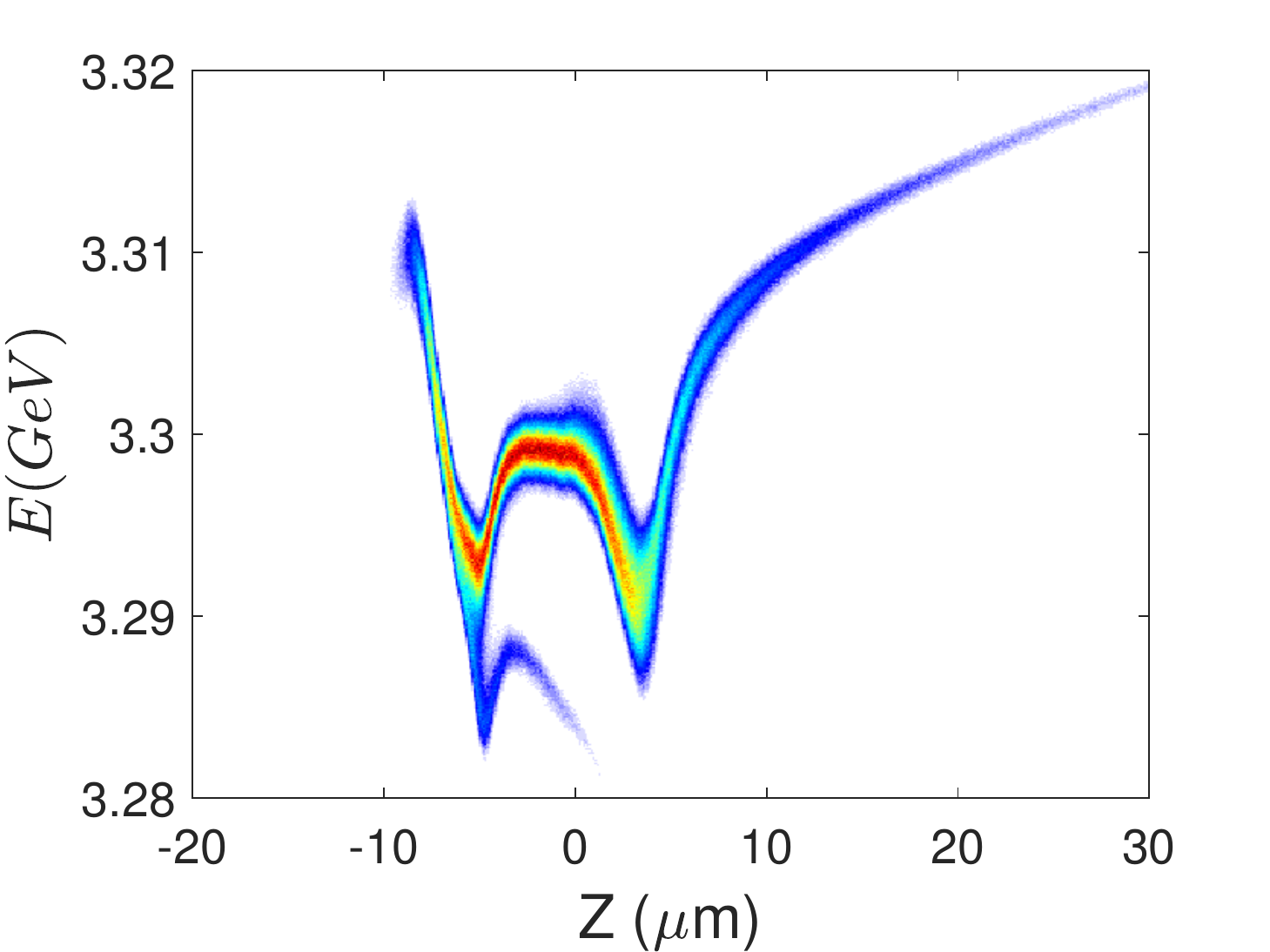}  
	\end{minipage}%
	\hfill 
	\begin{minipage}[htbp]{0.25\linewidth}  
		\centering  
		\includegraphics[width=\columnwidth]{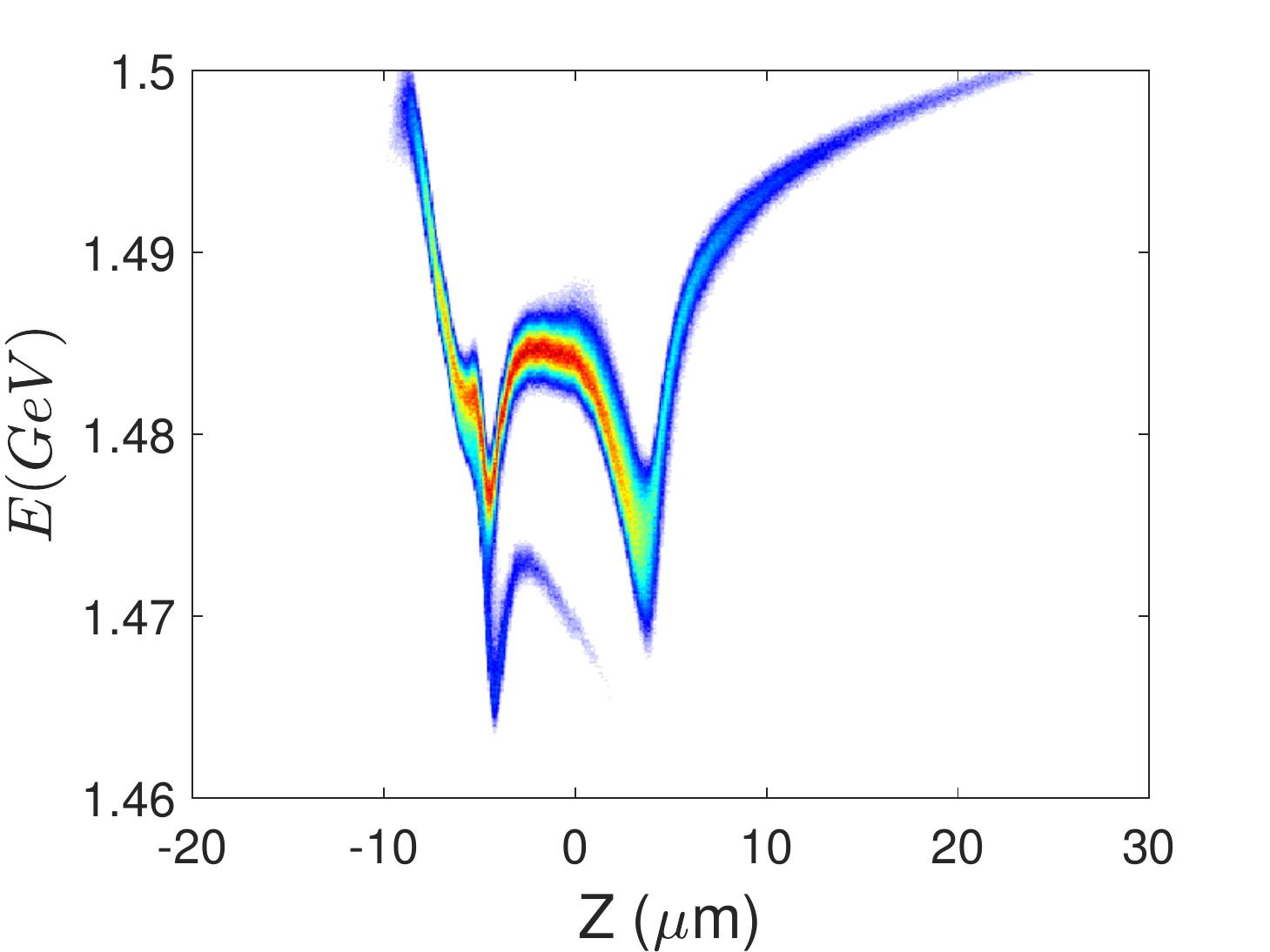}  
		
	\end{minipage}%
	\hfill 
	
	\begin{minipage}[htbp]{0.25\linewidth}  
		\centering  
		\includegraphics[width=\columnwidth]{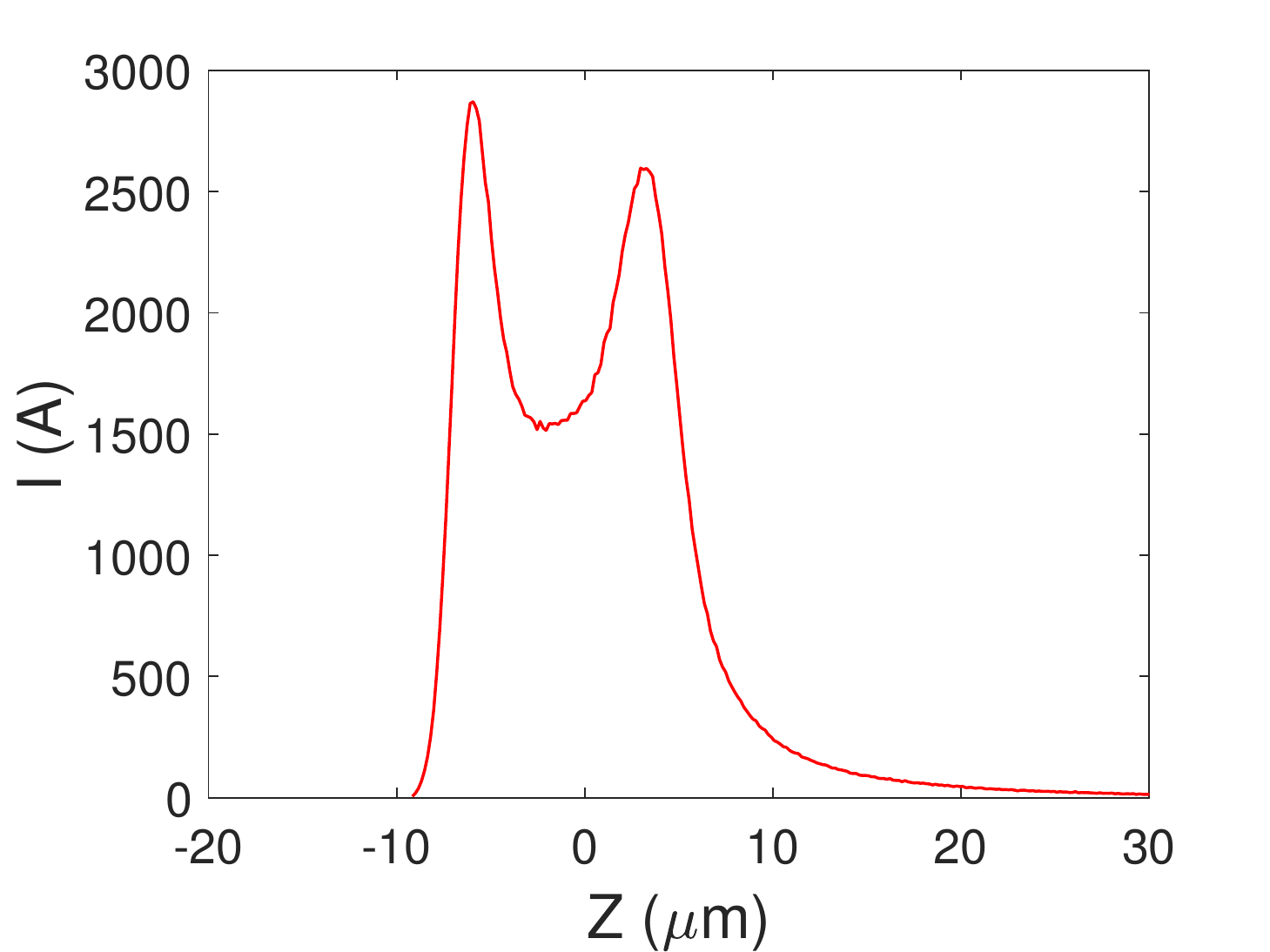}  
		\centerline{(a)}  
	\end{minipage}%
	\hfill 
	\begin{minipage}[htbp]{0.25\linewidth}  
		\centering  
		\includegraphics[width=\columnwidth]{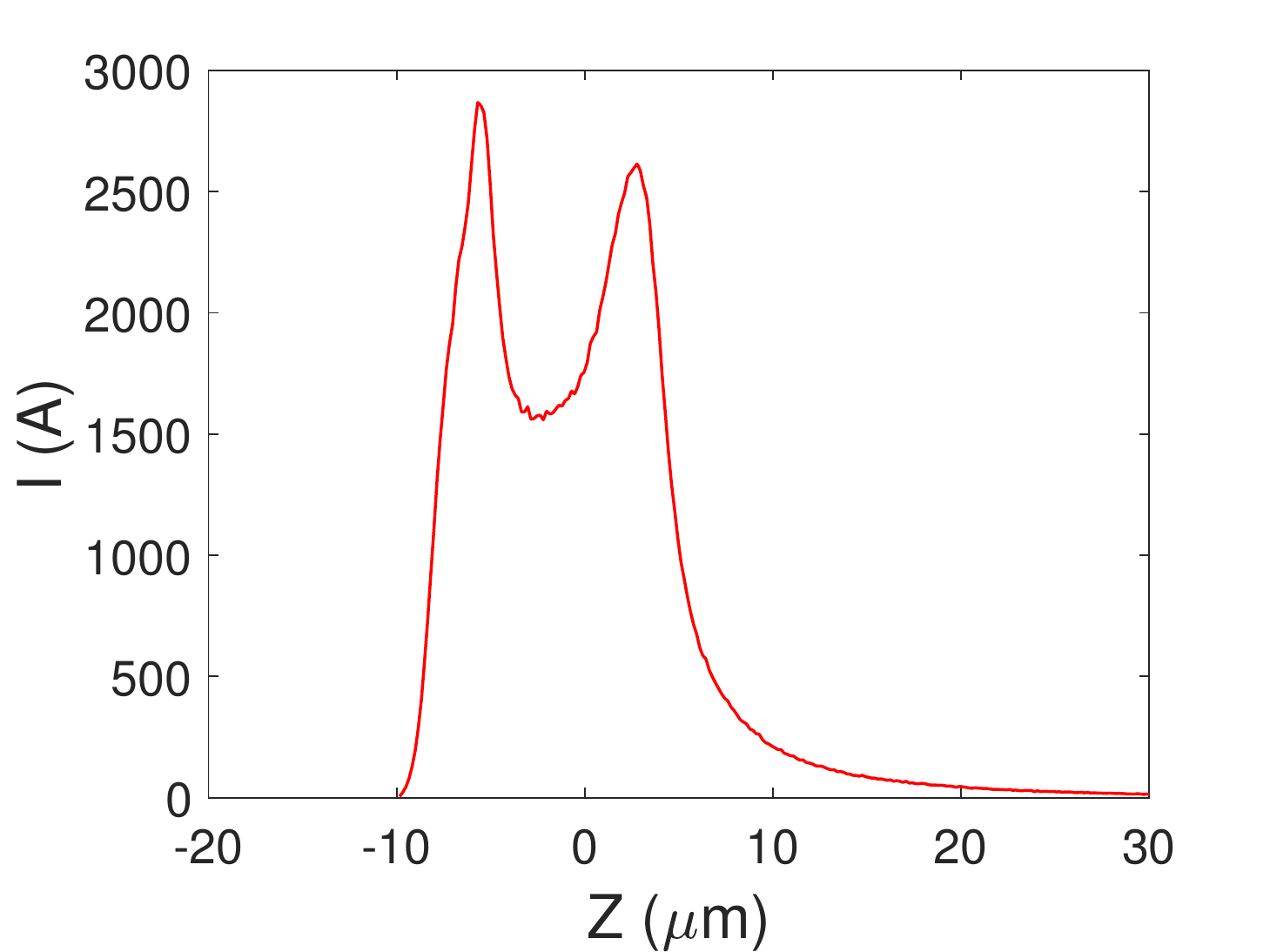}  
		\centerline{(b)}
	\end{minipage}%
	\hfill
	\begin{minipage}[htbp]{0.25\linewidth}  
		\centering  
		\includegraphics[width=\columnwidth]{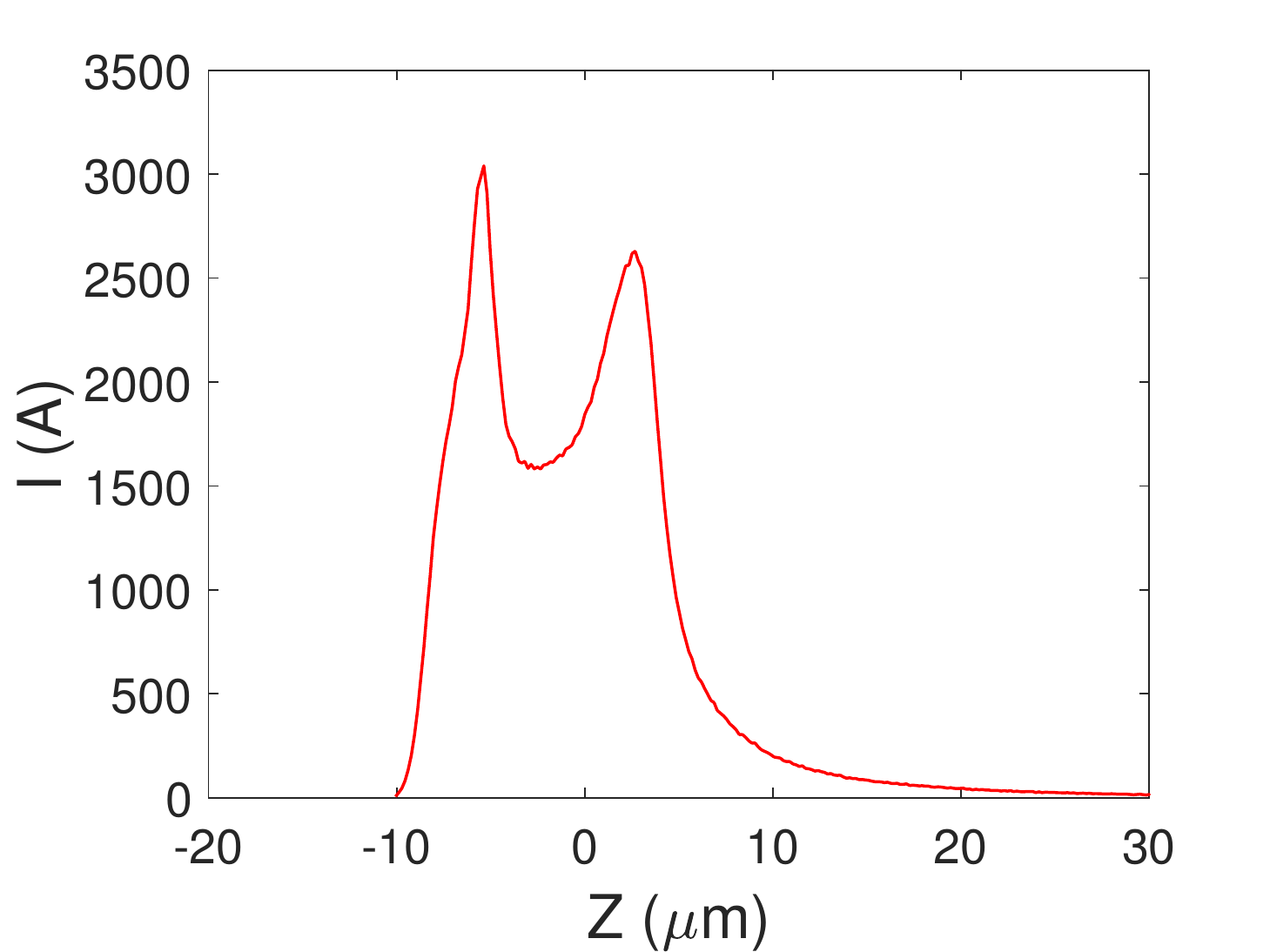}  
		\centerline{(c)}  
	\end{minipage}%
	\hfill 
	\begin{minipage}[htbp]{0.25\linewidth}  
		\centering  
		\includegraphics[width=\columnwidth]{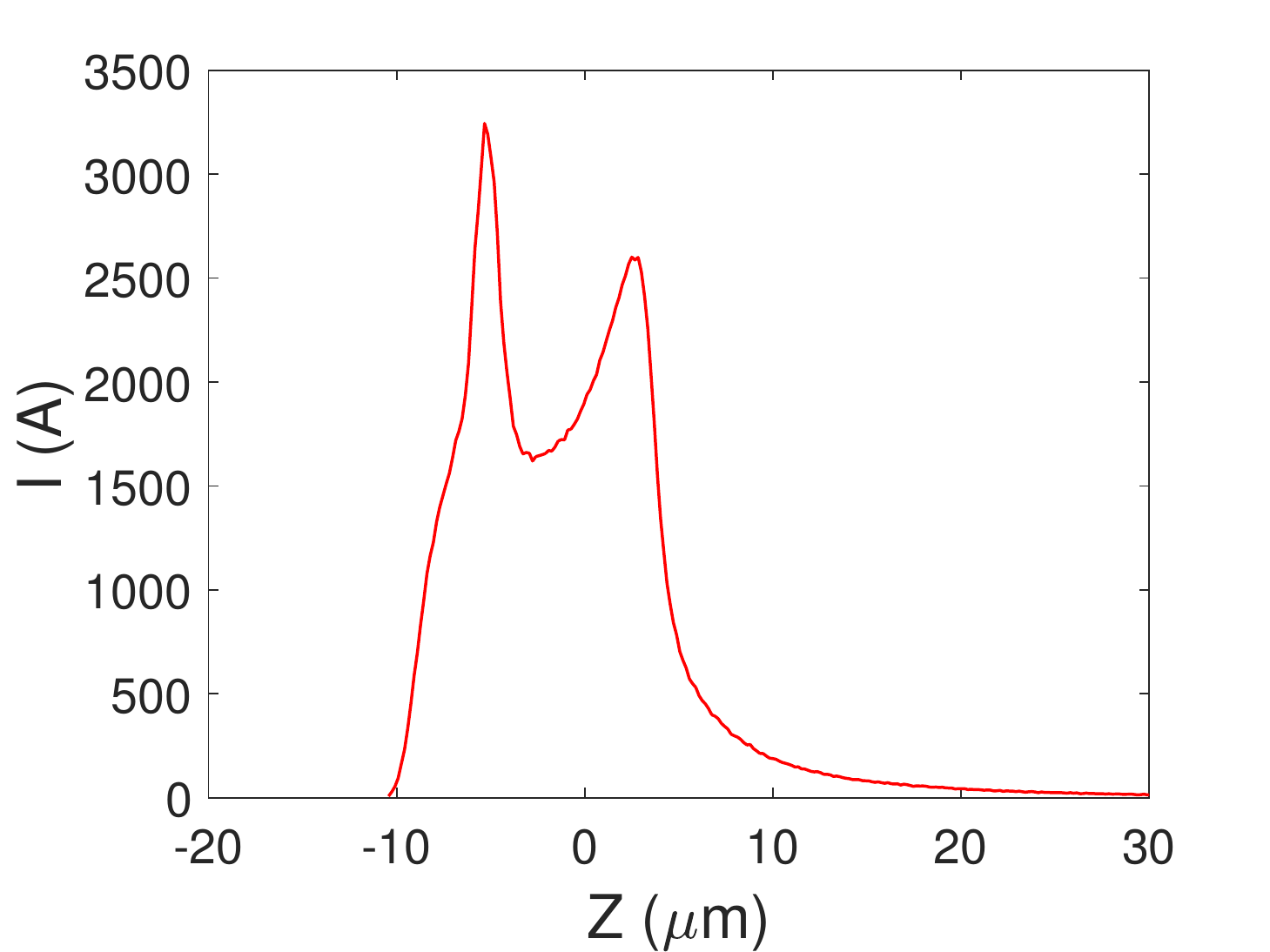}  
		\centerline{(d)}
	\end{minipage}%
	
	\caption{Longitudinal phase spaces (upper) and current profiles (lower) of the normally accelerated electron beam (a) and the three delayed electron beams with delay distances of 57.70 (b), 76.91 (c), and 115.41 mm (d) at the end of the linac. The beam’s head is to the left.}
	\label{FIG5} 
\end{figure*} 

\begin{table*}[!htb]
	\renewcommand\arraystretch{1.5}
	\centering
	\setlength{\tabcolsep}{3.5mm}
	\caption{\label{tabfourpara}%
		Parameters of the four typical electron beams and the corresponding spectrum ranges.
	}
	\begin{tabular}{cccccccc}
		\hline 
		\textrm{$\rm \theta (\degree)$}&
		\textrm{$\rm \theta_{m} (\degree)$}&
		\textrm{$\rm \Delta Z (mm)$}&
		\textrm{$\rm E (GeV)$}&
		\textrm{$\rm \epsilon_{nx} (mm\cdot mrad) $}&
		\textrm{FEL-I $\rm (keV)$}&
		\textrm{FEL-II $\rm (keV)$}&
		\textrm{FEL-III $\rm (keV)$}\\
		\hline 
		0     & 0    & 0  & 8.74 & 0.22          & 3.58-17.90 & 0.48-3.58 & 11.97-29.84\\ 
		1.993  & 0.130   & 57.70  & 5.11 & 0.24  & 1.22-6.12 & 0.16-1.22 & 4.09-10.20\\
		2.301  & 0.151   & 76.91  & 3.30 &  0.28   & 0.51-2.55 & 0.07-0.51 & 1.71-4.25\\
		2.819  & 0.184    & 115.41  & 1.48 & 0.33  & 0.10-0.51 & 0.01-0.10 & 0.34-0.86\\ 
		\hline
	\end{tabular}
\end{table*}

This delay system is designed to be placed before the L4 of the SHINE to achieve the multi-beam-energy operation. In the baseline design of the SHINE, the electron beam energy is about 5.11 GeV at the exit of the L3. In the L4, the energy difference between the delayed electron beams and the undelayed electron beams becomes larger and larger. For example, if the electron beams are delayed by half of the rf period of the L4, they will be decelerated to below 1.5 GeV at the exit of the L4 and those without delay will be accelerated to over 8 GeV. This will cause those electron beams with large energy differences to meet the same quadrupoles behind the delay system but with very different strength. The quadrupoles behind the delay system that are previously designed to control the normally accelerated electron bunches are difficult to simultaneously control those delayed electron bunches. Based on the previously designed quadrupoles, the electron bunches decelerated to below 1.5 GeV will have $\rm \beta$ functions greater than 100000 m at the end of the linac. Therefore, the strength of the quadrupoles in the L4 and the dechirper section needs to be redesigned. To control the envelopes of the electron bunches with large energy differences at the same time, the evolutionary algorithm is also used to optimize the 41 quadrupoles in the L4 and the dechirper section. The results show that the envelopes of the electron bunches whose energy is within the adjustment range of the delay system can be well controlled by the optimized lattice. Based on the optimized lattice, the $\rm \beta$ functions of four electron bunches with typical energies of 8.74, 5.11, 3.30, and 1.48 GeV, respectively, along the L4 and the dechirper section are shown in Fig.\ \ref{FIG4}. The corresponding delay lengths of the four electron bunches are 0, 57.70, 76.91, and 115.41 mm, respectively.

To further analysis the feasibility of the delay system, the start-to-end simulation of the SHINE containing the delay system is performed. Tracking simulation in the injector is performed by ASTRA \cite{astra} where transverse space charge forces are strong. ELEGANT is used to track electron beams in the main linac and the delay system where CSR effect, longitudinal space charge, and wakefields are considered. GENESIS \cite{genesis} is used to simulate the XFEL generation.

The longitudinal phase spaces and current profiles of the electron beam without delay and the three delayed electron beams at the end of the linac are shown in Fig.\ \ref{FIG5}. As shown in Fig.\ \ref{FIG5} (b, c, and d), the longitudinal phase spaces of the delayed electron beams are changed by the CSR effect in the delay system. Extra energy chirp is introduced at the head and tail of the electron bunches while the corresponding current profiles of the electron bunches are not greatly changed. Parameters of the four electron beams, including the corresponding dipole angle of the DBA, micro-bend angle, delay distance calculated by ELEGANT, energy and normalized horizontal emittance at the end of the linac and the adjustable spectrum ranges of the three undulator lines are shown in Tab.~\ref{tabfourpara}. The delay distance with a value of 0 refers to the undelayed electron beam. As the delay distance increases, the bend angle needs to be increased correspondingly, which leads to an increase in the emittance induced by the CSR effect. The emittance of the electron beam delayed by 115.41 mm has risen from 0.22 to 0.33 $\rm mm\cdot mrad$, which is still within the acceptable range. 

\begin{figure}  
	%
	%
	%
	\begin{minipage}[htbp]{0.5\linewidth}  
		\centering  
		\includegraphics[width=\columnwidth]{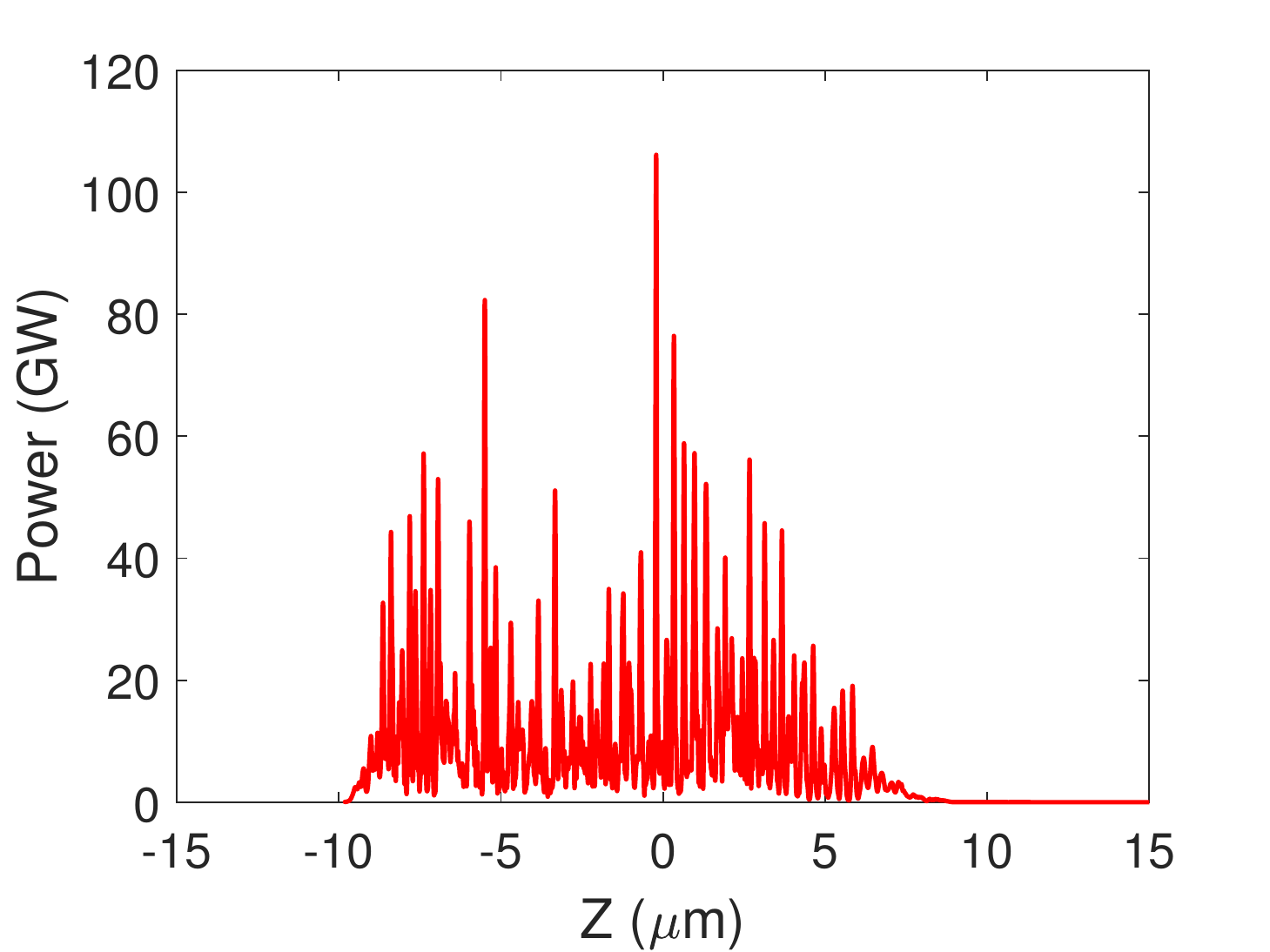}  
	\end{minipage}%
	\hfill 
	\begin{minipage}[htbp]{0.5\linewidth}  
		\centering  
		\includegraphics[width=\columnwidth]{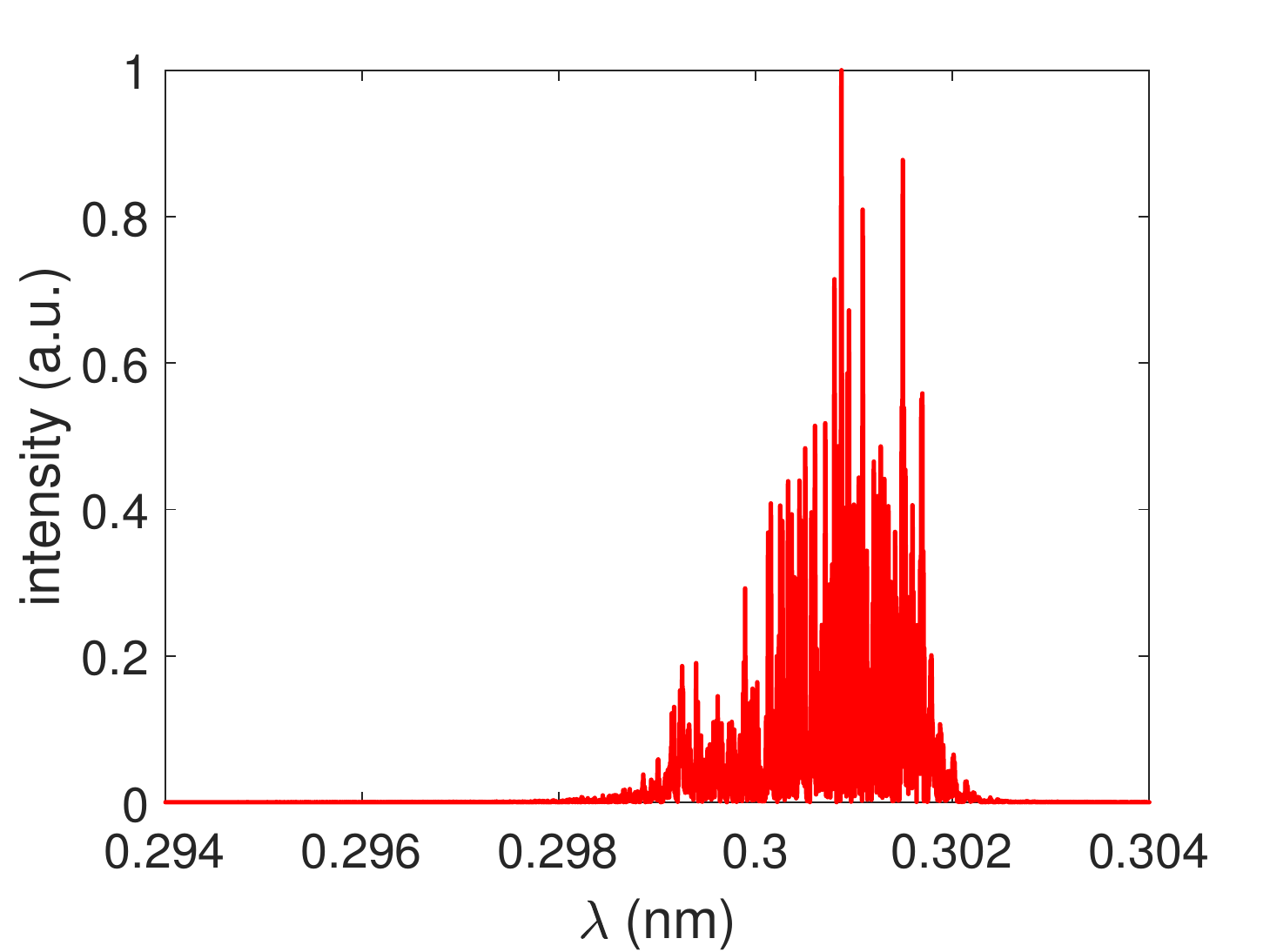}  
	\end{minipage}

	\caption{Simulated FEL power profile (left) and spectrum (right) of the delayed electron beam with an energy of 5.11 GeV. }
	\label{felresult} 
\end{figure}

The adjustable spectrum ranges of the three undulator lines based on the delayed electron bunches become much wider than their baseline designs. Taking the FEL-III as an example, since the strongest magnetic field strength of its undulator is 1.58 T, the longest radiation wavelength that can be generated based on the 8.74 GeV electron beam is 0.1 nm. Based on the delayed electron beam with an energy of 5.11 GeV, FEL-III can generate radiation with wavelengths up to 0.3 nm. The FEL power profile and spectrum obtained by the delayed electron beam with an energy of 5.11 GeV are shown in Fig.\ \ref{felresult}. There are 40 undulator segments of the FEL-III are used and appropriately tapered to maximize the pulse energy. The obtained pulse energy and rms bandwidth are 615 $\rm \mu J$ and $0.29 \%$, respectively. Radiation with longer wavelengths can be obtained with a lower energy electron beam. A 1.48 GeV electron beam can be used to generate XFEL pulses with wavelengths up to 3 nm.

\section{CONCLUSIONS AND OUTLOOK}

In this paper, a delay system based on four DBAs is proposed for the multi-beam-energy operation of the CW XFELs. The lattice of the delay system is designed to be achromatic and isochronous to preserve the electron beam quality. The optics balance method is adopted to cancel the CSR kick. Besides, the lattice behind the delay system is also optimized to simultaneously control the electron beams with large energy differences. Start-to-end simulations based on the SHINE parameters prove that the delay system can flexibly generate electron beams with energy between 1.48 to 8.74 GeV. 

The proposed method in this paper can obtain bunch-to-bunch energy changed electron beams, which enlarges the adjustable spectrum range and greatly improves the usability of XFEL. Compared to the branching-off design, the delay system saves a lot of components for electron beam transport from the break section to the undulator hall and in particular saves valuable accelerator tunnel spaces. Compared to the scheme of changing the trigger frequency of the rf units, this delay system is more suitable for those machines operated in CW mode. 

Besides the DBA, the delay system can also be composed of other kinds of lattice such as the triple bend achromat. The extra energy chirp caused by the CSR effect in the delay system will broaden the bandwidth of XFEL pulses, which needs further optimization. In addition to delaying electron beams, this delay system may be used to control the distance between two electron bunches in the twin bunch operation mode \cite{marinelli2015high} by using the rf kickers to separate the two bunches with very close spacing. Moreover, since two adjacent electron bunches can be respectively at the decelerating phase and the accelerating phase of the accelerating structure behind the delay system, this may allow the electron bunch at the accelerating phase to get more energy like an energy recovery linac, which requires further research.

\section*{Acknowledgments}
The author would like to thank S. Chen, D. Gu, N. Huang, M. Zhang, B. Liu, D. Wang, and Z. Zhao for helpful discussions on beam dynamics and the SHINE projects. This work was partially supported by the National Key Research and Development Program of China (2018YFE0103100, 2016YFA0401900) and the National Natural Science Foundation of China (11775293).


\bibliography{mybibfile}
\end{document}